# Latency Reduction for Mobile Backhaul by Pipelining LTE and DOCSIS

Jennifer Andreoli-Fang
Office of the CTO and R&D
CableLabs
Boulder, CO

John T. Chapman
Chief Technology and Architecture Office
Cisco Systems
San Jose, CA

*Abstract*—The small cell market has been growing. To backhaul wireless traffic from small cells, the mobile network operators (MNOs) are looking into economically viable solutions, specifically the hybrid fiber coaxial networks (HFC), in addition to the traditional choice of fiber. When the latencies from both the wireless and the HFC networks are added together, it can result in noticeable end-to-end system latency, particularly under network congestion. If the two networks could somehow coordinate with each other, it would be possible to decrease the total system latency and increase system performance. In this paper, we propose a method to improve upstream user-to-mobile core latency by coordinating the LTE and HFC scheduling. The method reduces the impact on system latency from the HFC network's request-grant-data loop, which is the main contributor of backhaul upstream latency. Through simulation, we show that coordinated scheduling improves overall system latency.

I. INTRODUCTION

The mobile network operators have been under pressure to deploy dense small cell networks in response to the tremendous growth in mobile data usage [1]. All this traffic needs to be backhauled to the mobile core. Fiber has been traditionally preferred by the MNOs to backhaul macrocells. However, fiber is sparse, and to install fiber ubiquitously to backhaul the dense small cell locations is not as economical as using existing fixed infrastructure. Cable operators have built and deployed HFC networks everywhere to service broadband residential and commercial customers. This HFC network is an attractive option for small cell backhaul as it is ubiquitous, has ample capacity, and can provide a lower cost alternative to running new fiber.

The HFC network may incur higher latency than the allocated timing budget for mobile backhaul. The Data Over Cable Service Interface Specifications (DOCSIS®) [2][3] defines the protocol that governs the communication of broadband data between the cable modems (CMs) and the cable modem termination system (CMTS) over the HFC network. The DOCSIS upstream today incurs a minimum latency of 5 ms, average of 11-15 ms, and can be 20 to 50 ms for a loaded network [4]. In comparison, the LTE backhaul budget is lower than the minimum DOCSIS latency [5]. Interference coordination techniques such as CoMP require 5 ms of X2 latency to realize significant gain [6][7]. Furthermore, most of 5G applications require 10 ms user-to-core latency, with some ultra-low latency applications requiring 1 ms latency [8][10]. These requirements are not possible to meet with DOCSIS backhaul today. Latency on the order of a couple of milliseconds is challenging even for a fiber backhaul network [9].

Both LTE and HFC networks follow a request-grant-data (REQ-GNT-data) loop for typical upstream transmissions. When wireless data is sent from the user equipment (UE) to the LTE base station (eNB) and backhauled over DOCSIS, the data experiences this type of 3-way loop twice. This occurs when the two networks, the wireless and its backhaul, are uncoordinated.

In this paper, we propose a method to improve the upstream UE-to-mobile core latency by coordinating the LTE and DOCSIS scheduling operations. A 2-stage pipeline is formed with the LTE and DOCSIS schedulers. The underlying DOCSIS scheduler is modified to be LTE aware. The end result is significantly reduced system latency with more deterministic jitter.

This paper is organized as follows. Section 2 provides a brief introduction on the DOCSIS and LTE upstream scheduling operations. Section 3 discusses the proposed method in detail. Section 4 describes an end-to-end MATLAB simulator environment including DOCSIS and LTE components developed to investigate the benefit of coordinated scheduling operations. Simulation results that show latency improvement are also described in the section.

II. BACKGROUND AND PREVIOUS WORK

*A. DOCSIS Upstream Data Plane Latency*

The DOCSIS protocol specifies several major types of upstream scheduling services including best effort (BE), real-time polling service (RTPS), and unsolicited grant service (UGS). Most upstream data is transmitted with BE service. Best effort scheduling follows a request-grant-data loop as shown in Fig. 1 (also see [4]). The requests are sent in the contention regions which the CMTS schedules regularly.

RTPS was designed to support real-time data flows that generate variable size packets periodically where the CMTS provides unicast request opportunities periodically. As shown in Fig. 1, after the CM detects data arrival and formulates a bandwidth request (REQ), it waits either for a contention region or a polling opportunity to transmit the REQ, depending on the scheduling service the traffic is configured for.

UGS was designed to support real-time data flows, such as VoIP, that periodically generate fixed size packets. The CMTS provides fixed-size grants of bandwidth on a periodic basis. The CM utilizes the periodic grants to transmit data directly without sending REQs.

The CMTS scheduler typically processes REQs every 2 ms and generates a MAP that describes 2 ms worth of grant allocations. Since the CMTS sends at least one MAP in advance of that MAP's allocation start time, the shortest REQ-GNT cycle on DOCSIS is theoretically 4 ms. Additionally, the CM and the CMTS each need processing and lead time, typically 0.5 ms on each device. So the practical minimum upstream latency for DOCSIS is about 5 ms.

The average latency is expected to be higher due to the random arrivals of data and therefore, REQs, as well as the scheduling of contention regions. Higher network loading may also result in collision of the REQs, which triggers a truncated exponential backoff that will further increase latency. The average DOCSIS upstream latency for best effort traffic has been measured to be 11-15 ms with the potential of a significantly higher maximum latency. For further discussion on this topic, see [4].

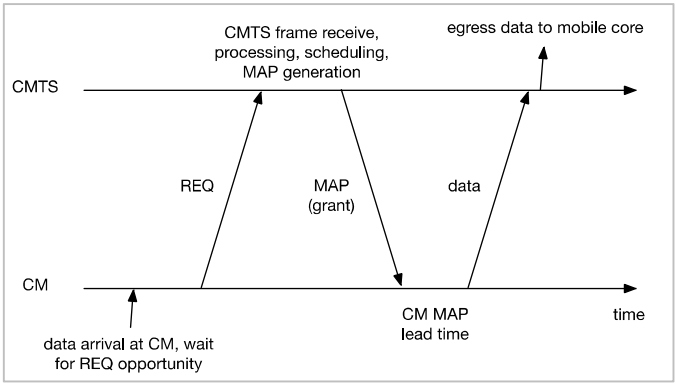

Fig. 1. DOCSIS REG-GNT-data loop for BE or RTPS scheduling service

### B. LTE Uplink Data Plane Latency

LTE uplink (UL) access follows a similar 3-way REQ-GNT-data loop but with some differences. Referring to Fig. 2a, which shows the signaling between UE and eNB, when data arrives at the UE, the UE first determines if it has a valid LTE UL grant. If it does not, the UE waits for a Scheduling Request (SR) opportunity.

Upon receiving the SR, the eNB schedules an UL grant for the UE to transmit a Buffer Status Report (BSR). The eNB turnaround time is 4 ms. Once the BSR is received by the eNB, it schedules an UL grant for the UE to transmit data. The typical minimum LTE uplink latency for transmitting one short packet with no network loading is summarized in Table I.

Note that the 21 ms average minimum latency for LTE is greater than the 5 ms minimum or 11-15 ms average latency for DOCSIS.

### C. Sum it Altogether

Referring again to Fig. 2a, when data arrives at the UE to be transferred uplink to the eNB, and backhauled on the DOCSIS link to the mobile core, it needs to traverse two 3-way REQ-GNT-data loops on LTE and on DOCSIS. First, the UE makes a scheduling request which is followed by an LTE grant that allows the data to traverse through the LTE air interface. Once the data reaches the eNB and is egressed to the CM, a DOCSIS request is made by the CM to the CMTS. The CMTS then assigns DOCSIS grant(s) which allow the data to traverse through the DOCSIS backhaul where it eventually reaches the mobile core. The LTE and DOCSIS latencies are added in serial.

TABLE I. LTE UPLINK LATENCY COMPONENTS

| Latency components | Latency (ms) |
|---|---|
| Waiting time for SR (assume configured SR period of 5 ms) | 0.5 – 5.5 |
| UE sends SR, eNB decodes SR, eNB generates grant for BSR | 4 |
| eNB sends grant, UE processes grant, UE generates BSR | 4 |
| eNB processes BSR, eNB generates grant for data | 4 |
| eNB sends grant, UE processes grant, UE sends UL data | 4 |
| eNB decodes UL data (estimate) | 1.5 – 2.5 |
| Total | 18 – 24 |

### D. Previous Work

To the best of our knowledge, there has been no previous attempt to reduce the latency by coordinating wireless and backhaul networks. Efforts have been made to separately reduce the LTE latency and the DOCSIS latency. The 3GPP is working on techniques including contention based data transmissions, and reducing the number of transmission time intervals [11]. However, the proposed techniques have yet to be standardized into the 3GPP Stage 2 specifications.

DOCSIS has two built-in mechanisms to address latency: UGS/RTPS and Active Queue Management (AQM). UGS cannot be used to backhaul LTE traffic, as UGS grants are fixed in size and packets cannot be fragmented. RTPS eliminates contention latency but does not address the fundamental issue of latency incurred with the REQ-GNT loop. AQM [12] reduces latency by selectively dropping packets once the transmit buffer exceeds a threshold. Besides the standardized mechanisms, the CMTS vendor Cisco has implemented a predictive MAC scheduler, which proactively grants in an intelligent manner [4].

The work described in this paper differ from all these efforts. Our method can be implemented in an API into the DOCSIS scheduler. This API allows the information on expected future data transmission from LTE to be sent to the DOCSIS scheduler. In this way, the method proposed here works with any existing CMTS schedulers and requires minimal change to the software.

### III. PIPELINING THE LTE AND DOCSIS SCHEDULING OPERATIONS

We assume one or multiple small cells are integrated with or attached to a cable modem (CM) via Ethernet. This section

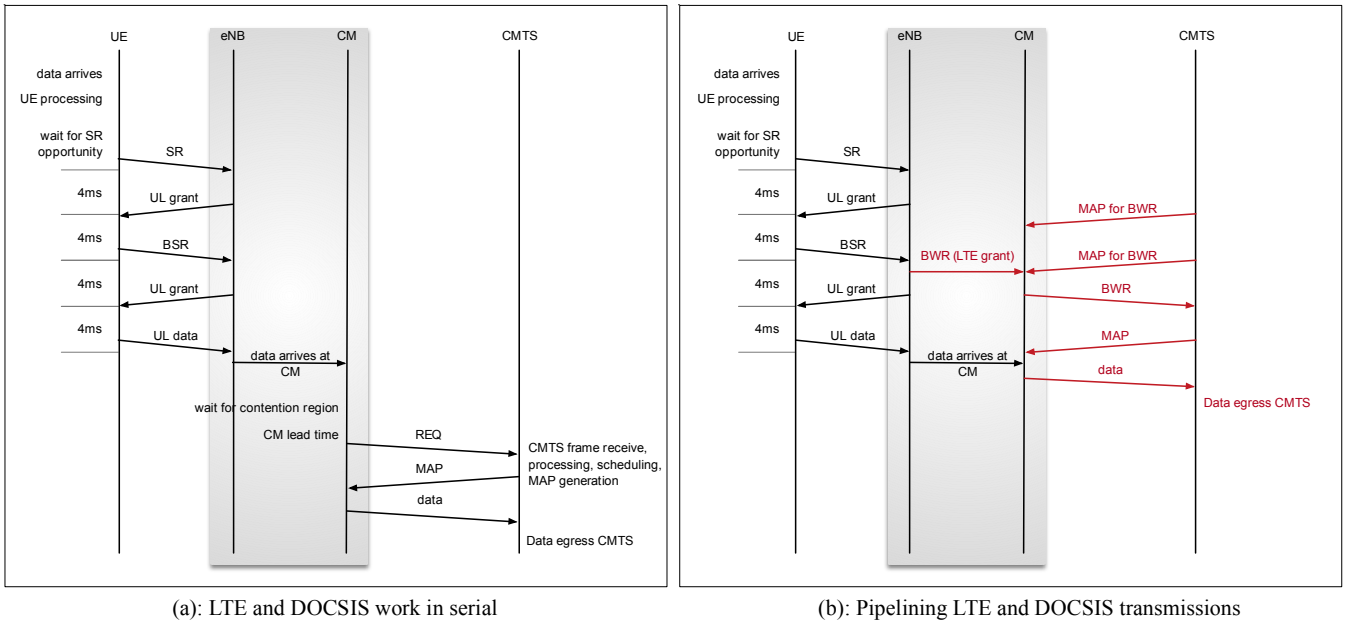

(a): LTE and DOCSIS work in serial  (b): Pipelining LTE and DOCSIS transmissions

Fig. 2. Backhauling LTE Data with a DOCSIS Network

proposes a method to improve the upstream latency by coordinating the LTE and DOCSIS scheduling loops. Since the DOCSIS REQ-GNT-data loop is the main contributor of backhaul upstream latency, the REQ-GNT-data processes on LTE and DOCSIS links are pipelined to reduce latency.

The first stage of the pipeline consists of the LTE REQ-GNT-data operation. Instead of waiting for the LTE data to arrive at the eNB, the eNB indicates to the CM that it is expecting UL data in the near future. This information prompts the second stage of the pipeline, the DOCSIS 3-way loop to start early. The DOCSIS scheduler prepares a grant in advance, just in time to transport the LTE data when it eventually arrives at the CM.

Effectively, the DOCSIS activities of sending a DOCSIS REQ on the CM, and receiving, scheduling, MAP generation activities on the CMTS occur in parallel to the activities taking place on the wireless link. In this way, the operation through the two networks become pipelined, where the first pipeline stage is the LTE scheduler, which informs the next stage, the DOCSIS scheduler, of what is about to come.

The details of the pipelining operations is shown in Fig. 2b. After the eNB has received the BSR and generated an LTE UL grant, the eNB MAC generates a Bandwidth Report (BWR) message, indicating the amount of grant and the grant time. This message is forwarded to the CM. Generally the eNB and the CM are connected via a Gigabit Ethernet connection and the propagation time of the message is negligible.

Meanwhile, on the DOCSIS side, the CMTS periodically allows the CM to transmit the BWR by sending a DOCSIS upstream (US) grant allocation MAP via the UGS scheduling service as described earlier in Section 2. Since the eNB scheduler generates UL data allocation every subframe, the granting interval on the DOCSIS link can also be every 1 ms. The CMTS scheduler uses the information included in the BWR message to schedule a transmission time for the UE data, and generates an US grant allocation MAP indicating the scheduled transmission start time to the CM. When LTE UL data egresses the eNB and arrives at the CM, a DOCSIS US grant has arrived and is waiting to be used at the CM to immediately forward the LTE data to the CMTS.

### A. The Bandwidth Report

The BWR needs to be sent to the CMTS quickly in order for the CMTS scheduler to make use of the time sensitive information and to generate an US grant. Therefore, the BWR is transported via the UGS scheduling service as stated previously, since it allows the CM to skip steps to request the bandwidth before sending the BWR.

The BWR can simply describe the LTE grant as a "bulk grant" in terms of a single block of bytes that the eNB MAC has allocated for transmission in future subframes. Since multiple UEs may be scheduled for transmission in a single UL subframe, the eNB MAC simply aggregates the individual LTE grants allocated for each UE in that subframe in a single block of bytes.

Alternatively, to allow for traffic differentiation on the DOCSIS link, the BWR can report the amount of UL grant for each Logical Channel Group (LCG). Each UE's BSR includes the amount of bytes requested for each of the UE's 4 LCGs. This information helps the eNB MAC scheduler determine the amount of bytes to be granted for each LCG, even though the UL grant is only expressed in terms of a single number of bytes. The BWR includes 4 blocks, with each block indicating the aggregate bytes allocated for an LCG for all the UEs scheduled to transmit in the future subframe described by the BWR. Because each LCG maps to a different set of QoS parameters, the CMTS scheduler uses this information to prioritize the LTE data traffic in a more granular manner.

The BWR must also indicate the expected egress time of the data it describes. This is generally 8 ms minus the amount of time it takes for the LTE scheduler to compute the LTE grant.

The BWR takes up a fraction of the DOCSIS upstream capacity. For an 80-byte length BWR, if a BWR is sent every 1 ms, then the BWR will consume 640 kbps per attached small cell.

*B. Effect of HARQ Failures*

The LTE UL employs 8 synchronous Hybrid Automatic Repeat reQuest (HARQ) processes to increase reliability and reduce latency associated with air interface failures. Retransmissions occur at fixed 8 ms intervals after the initial transmission has failed. For the BWR-based method, the eNB schedules the UL grant while also informing the DOCSIS link via BWR about the incoming LTE data that is expected to arrive at the CM 8-9 ms after the LTE grant. The CM forwards the BWR to requests resources from the CMTS for the expected LTE data.

If the initial transmission fails, the resources proactively granted on the DOCSIS side are wasted. The amount of unused DOCSIS grants depends on the HARQ failure rate. Let us define a given transport block size as $B$ and the total LTE resources to transmit the transport block as $G_{total}$. If the first transmission succeeds, $G_{total} = B$ and no grants are wasted on the DOCSIS link. With $k$ retransmissions,

$$G(k)_{total} = (k+1)B$$

The LTE grant utilization $G_{util}(N)$, as a function of number of retransmissions $N$, and a given target block-error-rate $BLER$, can be expressed as follows:

$$G_{util}(N) = \sum_{k=0}^{N} \frac{B}{G(k)_{total}} (1-BLER)(BLER)^k$$

$$= \sum_{k=0}^{N} \frac{1}{(k+1)} (1-BLER)(BLER)^k$$

The maximum number of HARQ retransmissions are configured by Radio Resource Control (RRC) signaling. It is 4 in a typical configuration. The expected grant utilization can be found by averaging up to the maximum number of retransmissions with $N = 4$, i.e., $E\{G_{util}\} = G_{util}(4) = 0.9482$, assuming a target BLER of 10% which is typical in a small cell deployment. Since there is a one-to-one relationship between the grant utilizations on the LTE link and the DOCSIS link due to HARQ retransmissions, the percentage of unused DOCSIS grants due to HARQ retransmissions is less than 6% for a typical HARQ operating point.

IV. SIMULATION AND RESULTS

In this section, we present results from a custom MATLAB simulator to analyze the latency performance of the baseline system and the proposed method. This unique MATLAB model contains models for the UE, the eNB, the CM and the CMTS.

Simulation assumptions are summarized in Table II. The simulation platform models the end-to-end LTE and DOCSIS system with protocol signaling capability relevant for this study. In the LTE stack, the Media Access Control (MAC) and Radio Link Control (RLC) layers are modeled in full with some PHY layer abstractions. In the simulation setup, multiple UEs are connected to a single eNB, and multiple UEs are capable of transmitting upstream traffic. Backhaul is provided by a CM communicating with a CMTS.

In this study, the BWR is constructed by the eNB every 2 ms. The message is then forwarded to the CM and is sent using the UGS scheduling service on the DOCSIS network. The frequency of the BWR transmission by the eNB is designed to match the periodicity of UGS grant as set by the CMTS which is 2 ms in this study. This can be reduced to 1 ms as needed. LTE data is backhauled on the DOCSIS network by using a best effort (BE) scheduling service for the baseline, and by using scheduled grants when the BWR is used.

TABLE II. SIMULATION SETTINGS

| Simulation parameters | |
|---|---|
| Simulation time | 2 seconds (2000 subframes) |
| **DOCSIS parameters** | |
| Number of CMTS and CMs | 1 CMTS, 1 CM |
| Scheduling service for BWR | UGS |
| BWR periodicity | 2 ms |
| UGS grant periodicity | 2 ms |
| Scheduling service for data | Best effort |
| **LTE system parameters** | |
| Number of eNBs | 1 or 4 |
| Number of UEs per eNB | 1 or 6 |
| Duplexing method | FDD |
| System bandwidth | 10 MHz |
| Spatial multiplexing | SISO |
| BSR periodicity | 10 ms |
| Channel | Slow fading with channel updates periodic of 10 ms (MCS 18-26) |
| HARQ | OFF or ON (10 % BLER) |
| **LTE eNB parameters** | |
| Scheduler type | Round robin |
| BWR periodicity | 2 ms |
| **Traffic parameters** | |
| Upstream traffic type | Case 1: VoIP (60 bytes @ 20 ms inter-arrival) Case 2: Facebook Live video streaming / upload |

The LTE deployment is FDD with a system bandwidth of 10 MHz for both uplink and downlink. It is assumed that the LTE system uses SISO (Single-Input Single-Output) transmission mode. The BSR is transmitted by the UE with a periodicity of 10 ms. HARQ processes are modeled for both bursty traffic such as live video streaming and periodic traffic such as in VoIP traffic. The LTE eNB employs a round-robin scheduler when multiple UEs are present in the system. A slow fading wireless channel model is assumed in which the channel stays static for the duration of 10 ms. The physical channel is modeled by varying a UE's MCS (modulation and coding scheme) such that the each UE's MCS has an average MCS index of 22 with a normal distribution between MCS indices 18 and 26.

In the simulations, we study 2 scenarios: 1) multiple UEs sending VoIP traffic on an unloaded upstream; 2) multiple UEs and eNBs performing Facebook Live video streaming upload.

For each scenario, we first consider the performance of the baseline scenarios where no coordination exists between the DOCSIS and the LTE networks. We then study the proposed BWR method where the eNB signals the expected future LTE transmission information to the DOCSIS system which enables just-in-time scheduling grants to be made by the CMTS. The latency of the end-to-end uplink transmission is measured in the simulations from the time a packet arrives at the UE to the time it egresses the network side of the CMTS. The measurement is sampled at the input and output of MAC layers in each side.

### A. Scenario 1

In this scenario, the CMTS serves a CM which is connected to an eNB. The eNB serves 6 UEs. Table III shows the DOCSIS and the system latency for UEs transmitting VoIP traffic using the baseline vs. the BWR-based method. Each entry of the table is averaged over the 6 UEs. We observe that the BWR-based method reduces the latency at a constant of 4 ms in this scenario.

The constant latency reduction is due to the deterministic nature of the VoIP traffic, where the VoIP packets are small enough compared to the capacity for each LTE subframe so that they are scheduled with no buffering or segmentation required at the UEs. The variations between the minimum and the maximum in the baseline run or the BWR run is due to HARQ retransmissions.

In this scenario, because the background traffic loading on the DOCSIS link is not modeled, the DOCSIS link incurs the theoretical minimum of 5 ms of latency, as discussed in Section 2. This is shown in Table III under DOCSIS baseline latency. The purpose of the BWR is to "hide" all or part of DOCSIS REQ-GNT cycle under the LTE latency. We observe that the BWR has eliminated all the possible DOCSIS REQ-GNT loop, which is 4 ms. What is left is the latency of the managed queue in the CM, which includes framing and error correction.

### B. Scenario 2

In this scenario, the CMTS is connected to a CM which backhauls traffic for 4 eNBs. Each eNB serves 6 UEs. Each eNB is configured with a separate DOCSIS best effort service flow for data transmission.

The goal of this scenario is to show the effectiveness of the BWR-based approach under more realistic traffic and channel loading conditions. A video trace was captured from a real Facebook Live session. Each UE randomly picks a starting point in the video trace and continuously loops through the trace, and performs live video streaming on the uplink. Based on the bitrate of the video trace, the aggregate upstream traffic from all eNBs is 31 Mbps on a DOCSIS link that has a capacity of 39 Mbps. Thus, the DOCSIS link is 80% loaded with upstream traffic. This configuration and traffic loading are used for both the baseline and BWR methods.

One of the 4 eNBs is selected to be the eNB under test (EUT) and implements the BWRs when evaluating the BWR-based method. The remaining 3 eNBs serve as background traffic generator and do not implement the BWR.

Fig. 3 shows the CDF for DOCSIS-only latency experience by the traffic from the EUT using the baseline and the BWR methods. Fig. 4 show the CDF for the combined DOCSIS and LTE system latency for the set of UEs attached to the EUT using the baseline and the BWR-based method.

In Fig. 3 and Table IV, we observe that the minimum DOCSIS-only latency is inline with the theoretical minimum DOCSIS latency as discussed in Section 2. As the traffic load is increased compared to scenario 1, the average and maximum latencies increased significantly as data packets contend for the limited DOCSIS channel resources.

In comparison, the BWR-based method yielded a DOCSIS-only minimum latency that is identical to what we observed in scenario 1. This means that for the minimum latency, the use of BWR has eliminated all the possible DOCSIS REQ-GNT loop latency which is 4 ms. The average and the maximum latencies are reduced significantly due to eliminating the contention and the scheduling delay experienced by the data packets on the DOCSIS link. The maximum latency of 16 ms observed in the BWR case is signficantly reduced from the maximum latency of 66 ms observed in the baseline case, which is due to the normal network congestion and queuing from the heavier traffic load.

In this more realistic environment, traffic loading is increased as a result of multiple eNBs sharing the same DOCSIS backhaul. When multiple service flows or CMs attempt to access the DOCSIS channel at the same time, the effect of the best effort contention and backoff, in addition to the normal network congestion and queuing can be seen in the form of higher latency. This effect increases the baseline latency, but does not affect the BWR-based method, since the BWR is sent using the UGS scheduling service, thus bypassing the contention. In this way, the gain from the BWR-based latency reduction method increases significantly in the more realistic environment. Because both the BWR and the data transmissions are scheduled separately on the DOCSIS network, i.e., the BWR and the data transmissions do not have to contend for the channel, the long tail average DOCSIS latency of 20 to 50 ms can be eliminated.

TABLE III. LATENCY (IN MILLISECONDS) FOR MULTIPLE UEs, VoIP TRAFFIC (SCENARIO 1)

|  | End to end (LTE+DOCSIS) | | | DOCSIS only | | |
|---|---|---|---|---|---|---|
|  | Min | Avg | Max | Min | Avg | Max |
| Baseline | 18.4 | 23.4 | 42.4 | 5.2 | 5.95 | 6.2 |
| BWR | 14.4 | 19.4 | 38.4 | 1.2 | 1.95 | 2.2 |

TABLE IV. LATENCY (IN MILLISECONDS) FOR MULTIPLE UEs, VIDEO STREAMING / UPLOAD (SCENARIO 2)

|  | End to end (LTE+DOCSIS) | | | DOCSIS only | | |
|---|---|---|---|---|---|---|
|  | Min | Avg | Max | Min | Avg | Max |
| Baseline | 17.40 | 38.05 | 96.20 | 5.2 | 16.97 | 66.20 |
| BWR | 13.40 | 25.54 | 56.20 | 1.2 | 3.46 | 16.20 |

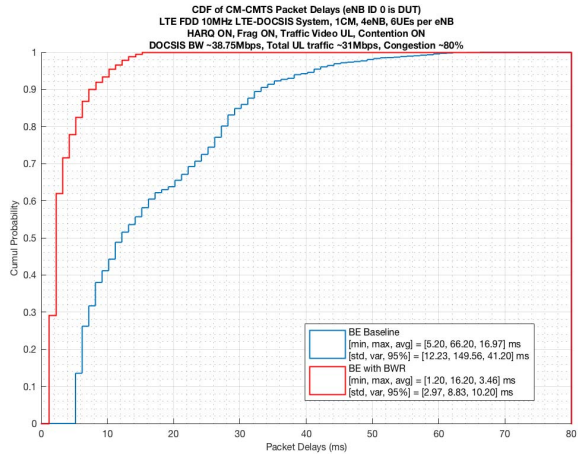

Fig 3. Latency CDF of the DOCSIS Link, Live Video Streaming / Upload

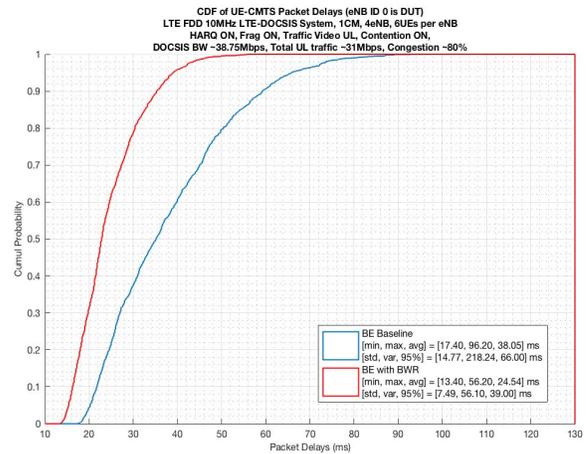

Fig 4. Latency CDF of the System, Live Video Streaming / Upload

## V. FUTURE WORK

The work presented so far focuses on standalone LTE small cells. Various standardization groups, such as the IEEE NGFI [13], the 3GPP [14] and the Small Cell Forum [15], have been working on a new centralized RAN architecture that splits the radio functions between baseband units and remote radio unit, stemming from MNO interests. The functional decompositions being studied range from splitting between the RLC and the PDCP layers, to intra-MAC layer split. Depending on how the LTE radio stack is split, the effect of HARQ failures and RLC layer segmentation will be different compared to that of a standalone LTE small cell.

## VI. CONCLUSION

In this paper, we present a method to improve the upstream system latency for a mobile network backhauled over a DOCSIS network. The operations of the two networks are pipelined. The first stage of the pipeline is the eNB scheduler: it constructs Bandwidth Reports based on its scheduling information. The BWR informs the second stage of the pipeline, the DOCSIS scheduler, of the traffic that will need to be backhauled in the near future. This enables the DOCSIS scheduler to proactively send just-in-time grants to the modem. Compared to non-coordinated system, the BWR-based method show sizeable latency reduction that becomes significant as the system becomes loaded.

## ACKNOWLEDGMENT

Many engineers provided technical insights throughout this project. The lead authors from CableLabs and Cisco would like to thank Joey Padden, Balkan Kecicioglu and Vaibhav Singh of CableLabs, and Oliver Bull, Alon Bernstein and Zheng Lu of Cisco for the technical discussions. The LTE and DOCSIS simulator environment was built by the following engineers: Michel Chauvin of CableLabs, and Elias Chavarria Reyes and Dantong Liu of Cisco.